\documentclass[12pt]{iopart}

\usepackage{iopams}  
\usepackage{graphicx}

\begin{document}
\title{A strange weak value in spontaneous pair productions via a supercritical step potential}

\author{Kazuhiro Yokota and Nobuyuki Imoto}
\address{Department of Materials Engineering Science,
Graduate School of Engineering Science, Osaka University,
Toyonaka, Osaka 560-8531, Japan}
\ead{yokota@qi.mp.es.osaka-u.ac.jp}

\date{\today}
\begin{abstract}
We consider a case where a weak value is introduced as a physical quantity rather than an average of weak measurements. The case we treat is a time evolution of a particle by 1+1 dimensional Dirac equation.
Particularly in a spontaneous pair production via a supercritical step potential, a quantitative explanation can be given by a weak value for the group velocity of the particle.
We also show the condition for the pair production (supercriticality) corresponds to the condition when the weak value takes a strange value (superluminal velocity).
\end{abstract}

\maketitle

\section{Introduction}
\label{sec:Int}
When a quantum system is prepared in an initial state $|\psi\rangle$ and then freely evolves, the expectation value of an observable $\hat{A}$ is calculated as $\langle\psi|\hat{A}|\psi\rangle$.
When, however, the system is prepared in $|\psi\rangle$ but postselected to be in a final state $|\phi\rangle$, the expectation value of the observable $\hat{A}$ is not given by $\langle\psi |\hat{A}|\psi\rangle$ but
\begin{eqnarray}
\langle\hat{A}\rangle_{\bf w}\equiv\frac{\langle\phi|\hat{A}|\psi\rangle}{\langle\phi|\psi\rangle}, \label{eq:wvdef}
\end{eqnarray}
which is referred to as ``weak value'' and was first introduced by Aharonov, Albert and Vaidman in 1988 \cite{W1}.
They also proposed how to experimentally obtain the weak value using weak measurement.
Weak measurement gives us a protocol to perform measurement without disturbance on a time evolution of a measured system.
The readout of such a measuring device (pointer) is noisy and gives us little information in a single run.
By averaging over many runs, however, we can correctly estimate the expectation value of measured observable, $\langle \psi|\hat{A}|\psi\rangle$, for the system in $|\psi\rangle$.
Especially, when the system is finally found in $|\phi\rangle$ (postselection), in the limit of no disturbance, the pointer shows the real part of the weak value, (\ref{eq:wvdef}).
Weak measurement provides an experimentally accessible manner to treat fundamental issues like quantum paradoxes \cite{Hardy1}-\cite{Hardy2}, and the applications for high sensitive measurement have been also reported \cite{sen1}-\cite{sen3}.
Using the weak value, we may consider a value of a physical quantity in the middle of a time evolution as in the case of classical physics.
Indeed, trajectories for an ensemble of quantum particles were recently observed via weak measurement \cite{traj}.
A wave function can be also directly observed as reported in \cite{wf, wf2}.
As shown in equation (\ref{eq:wvdef}), a weak value may take a strange value lying outside the eigenvalue spectrum of $\hat{A}$.
Hardy's paradox is a good example, in which we fall into a paradox when we consider which path a quantum particle takes.
Weak measurement actually results in the paradoxical values, and, at the same time, shows a strange weak value -1 for a projector unlike a probability \cite{Hardy1, Hardy2}.
Clarifying the condition for such a strange weak value may give us a clue to understand the significance of a weak value \cite{geo, str}.
Especially, it is notable that the correspondence of a strange weak value to a violation of Leggett-Garg inequality was pointed out \cite{LG}-\cite{LG2}.

As we have seen, a weak value, defined by equation (\ref{eq:wvdef}), come from weak measurement originally: A weak value is widely accepted as a result of measurement in trying to carry out measurement without disturbance, although it is often argued whether such a trial always results in the weak value \cite{wt1, wt2}.
The measurability has been certainly crucial to emphasize the significance of the weak value more than just a by-product from calculation.
However, a weak value, by itself, must be helpful for understanding or explaining a physical phenomenon in quantum mechanics, if the value is legitimate as a physical quantity.
In other words, weak measurement is a mere tool for extracting out a weak value.
Then, it should be asked whether the weak value can be found in quantum process irrespective of weak measurement and can act as a value of a physical quantity.

In this paper, motivated by this question, we pick up spontaneous pair productions via a supercritical step potential, in which a weak value of a group velocity gives us a quantitative explanation.
In our case, the weak value is introduced without measurement.
We also show that the condition for the pair production, namely, the supercriticality of the step potential corresponds to the appearance of a strange weak value.
In the next section, we show a case in which a weak value appears in a time evolution by $1+1$ dimensional space-time Dirac equation.
In section \ref{sec:pair}, we introduce a specific case of spontaneous pair productions via a supercritical step potential.
In section \ref{sec:main}, we show how a weak value is related to the pair production rate. We also consider the case of a strange value.
Section \ref{sec:con} is devoted to our conclusion.

\section{Weak value and Dirac equation}
\label{sec:W_D}
It is known that quantum random walk \cite{qrw1,qrw2} is useful on discussing Dirac equation \cite{Di1,Di2}.
The idea stems from so-called Feynman's checkerboard \cite{Ch1,Ch2}.
Let us consider $1+1$ dimensional space-time Dirac equation for free Hamiltonian.
The Hamiltonian $\hat{H}$ can be described with Pauli matrices $\hat{\sigma}_l$ as $\hat{H}=c\hat{\sigma}_z\hat{p}+mc^2\hat{\sigma}_x$, where $m$, $p$, and $c$ denote the mass, the momentum of the particle, and the velocity of light respectively.
With Trotter product formula, time evolution $U(t)$ can be derived from quantum random walk as follows,
\begin{eqnarray}
U(t) &=& [e^{-\frac{i}{\hbar}c\hat{\sigma}_z\hat{p}\epsilon}e^{-\frac{i}{\hbar}mc^2\hat{\sigma}_x\epsilon}]^{\frac{t}{\epsilon}} \label{eq:qwdi} \\
& \longrightarrow & \ e^{-\frac{i}{\hbar}(c\hat{\sigma}_z\hat{p}+mc^2\hat{\sigma}_x)t}=e^{-\frac{i}{\hbar}\hat{H}t} \ \ \ (\epsilon \ \longrightarrow \ 0),
\end{eqnarray}
where $\epsilon$ is a unit of time for quantum random walk.
In equation (\ref{eq:qwdi}), $e^{-\frac{i}{\hbar}mc^2\hat{\sigma}_x\epsilon}$ and $e^{-\frac{i}{\hbar}c\hat{\sigma}_z\hat{p}\epsilon}$ correspond to the operations of tossing a coin and shifting respectively.

Considering the above derivation, we can easily find another representation of the time evolution by the Dirac equation.
Let us expand equation (\ref{eq:qwdi}) in terms of $\hat{p}$ as follows,
\begin{eqnarray}
U(t) &=& [e^{-\frac{i}{\hbar}c\hat{\sigma}_z\hat{p}\epsilon}e^{-\frac{i}{\hbar}mc^2\hat{\sigma}_x\epsilon}]^{\frac{t}{\epsilon}} \nonumber \\
&=& [e^{-\frac{i}{\hbar}mc^2\hat{\sigma}_x\epsilon}]^{\frac{t}{\epsilon}}
-\frac{i}{\hbar}c\epsilon\sum_{\frac{t'}{\epsilon}=1}^{\frac{t}{\epsilon}}[e^{-\frac{i}{\hbar}mc^2\hat{\sigma}_x\epsilon}]^{\frac{t}{\epsilon}-\frac{t'}{\epsilon}}\hat{\sigma}_z[e^{-\frac{i}{\hbar}mc^2\hat{\sigma}_x\epsilon}]^{\frac{t'}{\epsilon}}\hat{p}+ \ \cdot\ \cdot \ \cdot \nonumber \\
&\longrightarrow& e^{-\frac{i}{\hbar}mc^2\hat{\sigma}_xt}{\cal T}\left[\exp\left(-\frac{i}{\hbar}c\int_0^tdt'\hat{\sigma}_z(t')\hat{p}\right)\right] \ \ \ (\epsilon \ \longrightarrow \ 0), \label{eq:app}
\end{eqnarray}
where $\hat{\sigma}_z(t)\equiv e^{\frac{i}{\hbar}mc^2\hat{\sigma}_xt}\hat{\sigma}_z e^{-\frac{i}{\hbar}mc^2\hat{\sigma}_xt}$, and ${\cal T}$ stands for a time ordering operator
\footnote{
As will be discussed later, we consider the Dirac equation when $t$ is very small.
Then, the limits of $\epsilon$ and $t$ must be kept in order as we perform.
The order of the limits in reverse will derive a different mathematical result.
}.
As we argue later, we consider an initial state of a plane wave solution like $e^{\frac{i}{\hbar}p_ix}|E_i\rangle \equiv \psi_{p_i}(x)|E_i\rangle$ with the energy $E_i$ and the momentum $p_i$, in which $|E_i\rangle$ has two components being independent of $x$.
The notation of $\psi(x)$ is used to show the energy propagation, namely, the group velocity explicitly.
$\psi_{p_i}(x)$ and $|E_i\rangle$ are respectively called the space part and the chirality hereafter.
If the chirality is finally found in $|E_f\rangle$ (postselection), apart from whether such a postselection is possible, the space part after a time evolution can be formally calculated as follows,
\begin{eqnarray}
& & \langle E_f|U(t)|E_i\rangle\psi_{p_i}(x) \nonumber \\
&=& \left[\langle E_f|e^{-\frac{i}{\hbar}mc^2\hat{\sigma}_xt}|E_i\rangle
-\frac{i}{\hbar}c\int_0^tdt' \langle E_f|e^{-\frac{i}{\hbar}mc^2\hat{\sigma}_xt}\hat{\sigma}_z(t')|E_i\rangle\hat{p}+ \ \cdot \ \cdot \right]\psi_{p_i}(x) \nonumber \\
&=& \sum_{n=0}^{\infty}\left(-\frac{ic}{n!\hbar}\right)^nF^{(n)}(t)\hat{p}^n\psi_{p_i}(x), \label{eq:ft_wv}
\end{eqnarray}
where
\begin{eqnarray}
F^{(n)}(t)=\left\{ \begin{array}{ll}
f_c^{(n)}(t)\langle E_f|\hat{\sigma}_z|E_i\rangle -f_s^{(n)}(t)\langle E_f|\hat{\sigma}_y|E_i\rangle & (n:{\rm odd}) \\
f_c^{(n)}(t)\langle E_f|E_i\rangle -if_s^{(n)}(t)\langle E_f|\hat{\sigma}_x|E_i\rangle & (n:{\rm even}) \\
\end{array} \right.
\end{eqnarray}
and, for $n\ge 1$,
\begin{eqnarray}
f_c^{(n)}(t)&=&\int_0^tdt_1\int_0^{t_1}dt_2 \ \cdot \ \cdot \ \cdot \int_0^{t_{n-1}}dt_n{\rm cos}\left[\frac{mc^2}{\hbar}(t+2\sum_{n'=1}^{n}(-1)^{n'}t_{n'})\right],
\end{eqnarray}
and,
\begin{eqnarray}
f_s^{(n)}(t)&=&\int_0^tdt_1\int_0^{t_1}dt_2 \ \cdot \ \cdot \ \cdot \int_0^{t_{n-1}}dt_n{\rm sin}\left[\frac{mc^2}{\hbar}(t+2\sum_{n'=1}^{n}(-1)^{n'}t_{n'})\right].
\end{eqnarray}
We define $f_c^{(0)}(t)={\rm cos}\frac{mc^2}{\hbar}t$ and $f_s^{(0)}(t)={\rm sin}\frac{mc^2}{\hbar}t$.
When the time $t$ is very small such that $\frac{mc^2}{\hbar}t\ll 1$, we can find $F^{(n)}(t)={\it O}(t^n)$ and the following approximation.
\begin{eqnarray}
& &\langle E_f|U(t)|E_i\rangle\psi_{p_i}(x) \nonumber \\ 
&=&\langle E_f|E_i\rangle
\left[1-\frac{i}{\hbar}mc^2\langle\hat{\sigma}_x\rangle_{\bf w}t-\frac{i}{\hbar}c\langle\hat{\sigma}_z \rangle_{\bf w}\hat{p}t+ \ {\it O}(t^2)\right]\psi_{p_i}(x) \nonumber \\
& \sim & \ \langle E_f|E_i\rangle e^{-\frac{i}{\hbar}mc^2\langle\hat{\sigma}_x\rangle_{\bf w}t} e^{-\frac{i}{\hbar}c\langle\hat{\sigma}_z \rangle_{\bf w}\hat{p}t}\psi_{p_i}(x)  \ \ \ (t \ \sim \ 0) \nonumber \\
& \sim & \ \langle E_f|E_i\rangle e^{-\frac{i}{\hbar}mc^2\langle\hat{\sigma}_x\rangle_{\bf w}t} \psi_{p_i}(x-c\langle\hat{\sigma}_z\rangle_{\bf w}t) \ \ \ (t \ \sim \ 0),
\end{eqnarray}
where $\langle \hat{\sigma}_l \rangle_{\bf w}$ is a weak value given by
\footnote{
As shown later, a weak value is given by only a real number as long as in our case.
For simplicity, we represent it without notating $Re$ explicitly.
Generally, a weak value has an imaginary part.
In the original paper of weak measurement \cite{W1}, the pointer is assumed as Gaussian function.
Then, the imaginary part brings about the momentum shift in contrast to the position shift by the real part.
The shift amount of the momentum is proportional to the variance of the pointer.
For a plane wave as we treat, such a momentum shift does not occur, because the space part is represented by $\delta$ function, namely, zero variance in $p$-representation.
}
\begin{eqnarray}
\langle \hat{\sigma}_l \rangle_{\bf w} &=& \frac{\langle E_f|\hat{\sigma}_l|E_i\rangle}{\langle E_f|E_i\rangle}. \label{eq:sigma_wv}
\end{eqnarray}
$\langle \hat{\sigma}_z\rangle$ corresponds to the group velocity in the unit of $c$ at $t=0$, as the average $c\langle\hat{\sigma}_z\rangle$ shows the group velocity conventionally.
The above discussion is similar to how weak measurement gives us a weak value in a pre-postselected system, although the position of the particle $x$ shifts like a pointer here \cite{superl1}-\cite{tun4}. 

For our later discussion, we also introduce the case with a potential $V(x)$ being dependent of only $x$.
In a similar way of equation (\ref{eq:app}), we can find the time evolution as follows, 
\begin{eqnarray}
U(t)&=& [e^{-\frac{i}{\hbar}c\hat{\sigma}_z\hat{p}\epsilon}e^{-\frac{i}{\hbar}mc^2\hat{\sigma}_x\epsilon}e^{-\frac{i}{\hbar}V(x)\epsilon}]^{\frac{t}{\epsilon}} \ \ \ (\epsilon \ \longrightarrow \ 0) \nonumber \\
&=& e^{-\frac{i}{\hbar}mc^2\hat{\sigma}_xt}e^{-\frac{i}{\hbar}V(x)t}{\cal T}\left[\exp\left(-\frac{i}{\hbar}c\int_0^tdt'\hat{\sigma}_z(t')\left(\hat{p}-\frac{\partial V(x)}{\partial x}t'\right)\right)\right]. \label{eq:appV}
\end{eqnarray}
Especially, when the potential is linear like $V(x)=\alpha x$ with the constant $\alpha$, equation (\ref{eq:appV}) can be described as follows,
\begin{eqnarray}
U(t) = e^{-\frac{i}{\hbar}mc^2\hat{\sigma}_xt}e^{-\frac{i}{\hbar}\alpha xt}&{\cal T}&\left[\exp\left(-\frac{i}{\hbar}c\int_0^tdt'\hat{\sigma}_z(t')(\hat{p}-\alpha t')\right)\right].  \label{eq:appE}
\end{eqnarray}
When the initial state in $\psi_{p_i}(x)|E_i\rangle$ is postselected by $|E_f\rangle$ on the chirality, the space part can be found as follows,
\begin{eqnarray}
& & \langle E_f|U(t)|E_i\rangle \psi_{p_i}(x) \nonumber \\
&=& \langle E_f|e^{-\frac{i}{\hbar}mc^2\hat{\sigma}_xt}e^{-\frac{i}{\hbar}\alpha xt}{\cal T}\left[\exp\left(-\frac{i}{\hbar}c\int_0^tdt'\hat{\sigma}_z(t')(\hat{p}-\alpha t')\right)\right]|E_i\rangle\psi_{p_i}(x) \nonumber \\
&=&\langle E_f|E_i\rangle e^{-\frac{i}{\hbar}\alpha xt}
\left[1-\frac{i}{\hbar}mc^2\langle\hat{\sigma}_x\rangle_{\bf w}t-\frac{i}{\hbar}c\langle\hat{\sigma}_z \rangle_{\bf w}\hat{p}t+ \ {\it O}(t^2)\right]\psi_{p_i}(x) \label{eq:appE_wv0} \\
&\sim & \langle E_f|E_i\rangle e^{-\frac{i}{\hbar}\alpha xt} e^{-\frac{i}{\hbar}mc^2\langle\hat{\sigma}_x\rangle_{\bf w}t} e^{-\frac{i}{\hbar}c\langle\hat{\sigma}_z\rangle_{\bf w}\hat{p}t}\psi_{p_i}(x) \ \ \ (t \ \sim \ 0)\nonumber \\
&\sim & \langle E_f|E_i\rangle e^{-\frac{i}{\hbar}mc^2\langle\hat{\sigma}_x\rangle_{\bf w}t} \psi_{p_i-\alpha t}(x-c\langle\hat{\sigma}_z\rangle_{\bf w}t) \ \ \ (t \ \sim \ 0). \label{eq:appE_wv}
\end{eqnarray}
Of course, we are not allowed to perform such a postselection arbitrarily: The above discussion is just an armchair theory at this time.
However, we found the case where the above weak value formalism is available.

\section{Spontaneous pair productions via a supercritical step potential}
\label{sec:pair}
\begin{figure}
  \begin{center}
	 \includegraphics[scale=0.5]{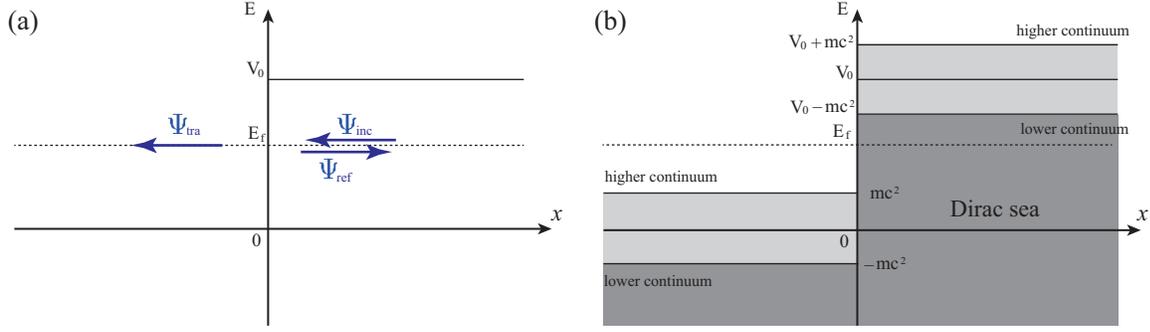}
  \end{center}
  \caption{(a)A step potential $V(x)=V_0\theta(x)$ with the step function $\theta(x)=1$ $(x>0)$, $0$ $(x<0)$.
(b)The energy spectrum with the step potential. The deep gray zone represents Dirac sea, in which particles are filled up.
The pale gray zone shows the forbidden energy levels. We call energy levels being higher (lower) than the maximum (minimum) forbidden energy level by higher (lower) continuum.}
  \label{fig:step}
\end{figure}

In this section, we review spontaneous pair productions via a supercritical step potential, which can be an application of equation (\ref{eq:appE_wv}).
Figure \ref{fig:step}(a) represents a step potential.
As shown in figure \ref{fig:step}(b), an energy level $E_f(>mc^2)$ sinks in the Dirac sea for $x>0$, if the potential height $V_0$ satisfies $V_0-mc^2>E_f$ (supercriticality).
An incident wave, $\Psi_{\rm inc}$, which comes from $x=\infty$, can be considered in such an energy level.
Passing $x=0$, the wave can transmit to $x=-\infty$ and result in a transmitted wave $\Psi_{\rm tra}$.
This is different from the case for so-called Klein's paradox, in which an incident wave comes from $x=-\infty$ to $x=0$ \cite{Kl1}-\cite{qft}.
A lack of a particle, namely, a hole in Dirac sea can be observed as an anti-particle.
Then, the transmission corresponds to a pair production of a particle and its anti-particle: In $x<0$, the transmitted flow represents the flow of produced particles. On the other hand, the net flow in $x>0$ can be regarded as the reversed flow of produced anti-particles, because the flow of particles in $x>0$ can be interpreted as the flow of particles to fill holes in the Dirac sea of $x>0$.
The transmitted particle number can be estimated in the same manner as a tunneling current \cite{tun}.
The number of particles arriving at $x=0$ per unit time can be given by $f(E(k))v(k)dk/(2\pi)$ within wavenumber $k\sim k+dk$.
$v(k)$ and $f(E(k))$ denote the group velocity and the Fermi distribution function for $x>0$.
With the transmission probability $T(k)$, the transmitted particle number, $N$, per unit time (pair production rate) can be estimated as follows,
\begin{eqnarray}
\frac{dN}{dt }&=& \int^{\infty}_{0}f(E(k))v(k)T(k)\frac{dk}{2\pi} \nonumber \\
&=& \frac{1}{2\pi\hbar}\int^{V_0-mc^2}_{mc^2}T(E)dE, \label{eq:rate}
\end{eqnarray}
where we have used $dE/dk=\hbar v(k)$ and the condition that $f(E(k))=1 \ (E\le V_0-mc^2)$, $0 \ (E>V_0-mc^2)$.
We have also taken account of no transmission for $E\le mc^2$ due to the Dirac sea and the forbidden energy levels.
In fact, quantum field theory also gives us the same result \cite{qft}.

The transmission probability $T(E)$ can be easily obtained.
As shown in figure \ref{fig:step}(a), the incident, reflected, and transmitted plane waves have the kinetic energies and the momentums 
($E_i, p_i$), ($E_i, -p_i$), and ($E_f, -p_f$) respectively.
We have defined them such that $p_i,p_f>0$ and $E_i\equiv E_f-V_0$, which satisfies $E_i<0$ in this case.
With coefficients $A$, $B$, and $D$, they can be described respectively as follows,
\begin{eqnarray}
\Psi_{\rm inc}(x) &=& \frac{A}{\sqrt{n_i^+}}e^{\frac{i}{\hbar}p_ix}
\left[ 
\begin{array}{c}
mc^2 \\
E_i-p_ic\\
\end{array}
\right] \ \equiv \ Ae^{\frac{i}{\hbar}p_ix}|E_i^+\rangle \\
\Psi_{\rm ref}(x) &=& \frac{B}{\sqrt{n_i^-}}e^{-\frac{i}{\hbar}p_ix}
\left[ 
\begin{array}{c}
mc^2\\
E_i+p_ic\\
\end{array}
\right] \ \equiv \ Be^{-\frac{i}{\hbar}p_ix}|E_i^-\rangle \\
\Psi_{\rm tra}(x) &=& \frac{D}{\sqrt{n_f^-}}e^{-\frac{i}{\hbar}p_fx}
\left[ 
\begin{array}{c}
mc^2\\
E_f+p_fc\\
\end{array} 
\right] \ \equiv \ De^{-\frac{i}{\hbar}p_fx}|E_f^-\rangle,
\end{eqnarray}
where $n_i^+$, $n_i^-$, and $n_f^-$ represent the normalization constants for the chiralities.
The energies and the momentums satisfy $E_f^2=(p_fc)^2+(mc^2)^2$ and $E_i^2=(p_ic)^2+(mc^2)^2$.
The group velocities of these waves can be given by $c\langle E_i^+| \hat{\sigma}_z| E_i^+\rangle=p_ic^2/E_i<0$, $c\langle E_i^-| \hat{\sigma}_z| E_i^-\rangle=-p_ic^2/E_i>0$, and $c\langle E_f^-| \hat{\sigma}_z| E_f^-\rangle=-p_fc^2/E_f<0$ respectively.
Although we cannot determine the coefficients, we can estimate the ratios of them from $\Psi_{\rm inc}(0)+\Psi_{\rm ref}(0)=\Psi_{\rm tra}(0)$.
This is enough to obtain the ratios of the probability currents (fluxes) $j=c\Psi^{\dagger}\hat{\sigma}_z\Psi$ \cite{D_eq}.
As a result, the transmission coefficient $T(E_f)$ and the reflection coefficient $R(E_f)$ can be given as follows,
\begin{eqnarray}
T(E_f)&=&\frac{j_{\rm tra}}{j_{\rm inc}}=\frac{4r}{(1+r)^2} \label{eq:trans} \\
R(E_f)&=&\frac{j_{\rm ref}}{j_{\rm inc}}=\frac{(1-r)^2}{(1+r)^2}(=1-T(E_f)),
\end{eqnarray}
with
\begin{eqnarray}
r=-\frac{p_i}{p_f}\frac{E_f+mc^2}{E_i+mc^2}.
\end{eqnarray}
They satisfy $0< T(E_f)\le 1$ and $0\le R(E_f)< 1$ due to $r>0$.

\section{A weak value in the spontaneous pair production}
\label{sec:main}
\begin{figure}
 \begin{center}
  \includegraphics[width=0.6\linewidth]{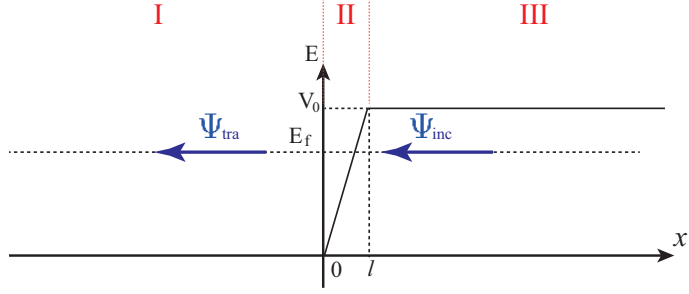}
  \caption{A model for a spontaneous pair production via a supercritical step potential.}
  \label{fig:step_wm}
 \end{center}
\end{figure}

We reconsider the pair production (transmission) process via the supercritical step potential in the context of a weak value discussed in section \ref{sec:W_D}. 
Let us begin with the potential as shown in figure \ref{fig:step_wm}, in which the sharp rise of the step potential at $x=0$ in figure \ref{fig:step}(a) is relaxed by the linear potential $V(x)=V_0x/l$ in the very narrow width $l$ (region ${\rm I\hspace{-.1em}I}$).
The incident wave in the energy level $E_f$ (with the kinetic energy $E_i=E_f-V_0$),
\begin{eqnarray}
\Psi_{\rm inc}(x) = e^{\frac{i}{\hbar}p_ix}|E_i^+\rangle = \psi_{p_i}(x) |E_i^+\rangle,
\end{eqnarray}
arrives at $x=l$ from $x=\infty$.
Let $t$ be the spending time in region ${\rm I\hspace{-.1em}I}$.
Due to the linear potential, the incident wave evolves by (see equation (\ref{eq:appE})),
\begin{eqnarray}
U(t)=e^{-\frac{i}{\hbar}mc^2\hat{\sigma}_xt}e^{-\frac{i}{\hbar}\frac{V_0}{l}xt}{\cal T}\left[\exp\left(-\frac{i}{\hbar}c\int_0^tdt'\hat{\sigma}_z(t')\left(\hat{p}-\frac{V_0}{l}t'\right)\right)\right]. \label{eq:appEE}
\end{eqnarray}
Because the transmitted wave can be described by $e^{-\frac{i}{\hbar}p_fx}|E_f^-\rangle\equiv\psi_{-p_f}(x)|E_f^-\rangle$, the chirality must be postselected by $|E_f^-\rangle$ in achieving the transmission at least.
Then, the space part is given by
\begin{eqnarray}
& & \langle E_f^-|U(t)|E_i^+\rangle \psi_{p_i}(x) \ \sim \ \psi_{p_i-\frac{V_0}{l}t}(x-c\langle\hat{\sigma}_z\rangle_{\bf w}^{\rm I\hspace{-.1em}I}t) \ \ \ (t \ \sim \ 0), \label{eq:wvtrans}
\end{eqnarray}
with the weak value,
\begin{eqnarray}
\langle \hat{\sigma}_z \rangle_{\bf w}^{\rm I\hspace{-.1em}I} &=& \frac{\langle E_f^-|\hat{\sigma}_z|E_i^+\rangle}{\langle E_f^-|E_i^+\rangle} \nonumber \\
&=& \frac{(E_i-E_f)+(p_i-p_f)c}{(E_i+E_f)+(p_i+p_f)c} = \frac{E_i-E_f}{(p_i+p_f)c}, \label{eq:wv_vel}
\end{eqnarray}
where, in r.h.s of (\ref{eq:wvtrans}), we have not cared the normalization and the phase.
We have assumed $t$ is very small at the moment.

Now, let us verify that this weak value formalism gives us a reasonable explanation for the spontaneous pair production.
In discussing a spending time for a tunneling particle in a barrier region, a problem often begins with how to define such a time in quantum mechanics.
Describing the time evolution with a weak value, we can provide a solution to the problem and explain seemingly superluminal tunneling.
It is similar to our discussion in section \ref{sec:W_D} \cite{tun3, tun4, tun1} (see also \cite{spl1, spl2}).
With a weak value, we can try to treat a value of a physical quantity in effect like classical mechanics.

In our case, what to discuss is whether we can work it out by thinking of an incident particle passing the region ${\rm I\hspace{-.1em}I}$ with the velocity $c\langle\hat{\sigma}_z\rangle_{\bf w}^{\rm I\hspace{-.1em}I}$ and the spending time $t$ as if it were a classical particle.
In the region ${\rm I\hspace{-.1em}I}$, the force $-V_0/l$ acts on a particle for time $t$, by which the momentum of the particle is changed as follows,
\begin{eqnarray}
\delta p =p_f'-p_i=-\frac{V_0}{l}t,
\end{eqnarray}
where $p_f'$ denotes the momentum after the impulse. Stated differently, we can find
\begin{eqnarray}
\frac{-l}{t}=\frac{V_0}{p_f'-p_i}=\frac{E_i-E_f}{p_i-p_f'}. \label{eq:vel_mo}
\end{eqnarray}
If $t$ represents the spending time in the region ${\rm I\hspace{-.1em}I}$, equation (\ref{eq:vel_mo}) shows the average velocity in the region ${\rm I\hspace{-.1em}I}$.
When the final momentum $p_f'$ is $-p_f$, equation (\ref{eq:vel_mo}) is identical to the velocity $c\langle\hat{\sigma}_z\rangle_{\bf w}^{\rm I\hspace{-.1em}I}$. 
Then, with the postselction $|E_f^-\rangle$ on the chirality, the time evolution (\ref{eq:appEE}) achieves the transition from $\psi_{p_i}(x)|E_i^+\rangle$ to $\psi_{-p_f}(x)|E_f^-\rangle$, which means $c\langle\hat{\sigma}_z\rangle_{\bf w}^{\rm I\hspace{-.1em}I}$ is the requisite velocity to achieve the momentum transition from $p_i$ to $-p_f$.
Due to $l \ \longrightarrow \ 0$ for the step potential, $t$ should finally satisfy $t \ \longrightarrow \ 0$ to converge the value (\ref{eq:vel_mo}), which makes the assumption that $t$ is very small in deriving equation (\ref{eq:wvtrans}) appropriate
\footnote{
Strictly speaking, we need an additional term being proportional to $i\alpha t^2\sim O(t)$ for the exponential approximation in equation (\ref{eq:appE_wv0}) due to $\alpha = V_0/l\sim O(t^{-1})$. However, this makes a change to just the phase and does not affect the group velocity.
}.

In fact, we can also estimate the pair production rate with the weak value, $c\langle\hat{\sigma}_z\rangle_{\bf w}^{\rm I\hspace{-.1em}I}$, for the velocity in the region ${\rm I\hspace{-.1em}I}$ (and furthermore at $x=0$ in figure \ref{fig:step}(a)).
As shown in equation (\ref{eq:rate}), all that is required is the transmission probability $T(E)$, which gives us the pair production probability within the energy spectrum $E\sim E+dE$, where a probability is concerned about one particle arriving at $x=0$ from $x=\infty$.
The reflection probability $R(E)=1-T(E)$ can be regarded as the probability for holding the vacuum state.
Now, we consider the fluxes for such probabilities in the regions I, ${\rm I\hspace{-.1em}I}$ ($x=0$), and ${\rm I\hspace{-.1em}I\hspace{-.1em}I}$.
In the region I, the flux $j=j_{\rm tra}$ can be described by $j=\rho_f c\langle E_f^-|\hat{\sigma}_z|E_f^-\rangle$, where $\rho_f$ stands for the probability density and $c\langle E_f^-|\hat{\sigma}_z|E_f^-\rangle$ is the velocity.
On the other hand, the net flux in the region ${\rm I\hspace{-.1em}I\hspace{-.1em}I}$, which can be represented by $\rho_i c\langle E_i^+|\hat{\sigma}_z|E_i^+\rangle$ with the probability density $\rho_i$, is also given by $j$ ($=j_{\rm inc}-j_{\rm ref}$).
Then, the average velocity of the incident, transmitted and reflected waves for one particle can be given by $(\rho_i c\langle E_i^+|\hat{\sigma}_z|E_i^+\rangle + \rho_f c\langle E_f^-|\hat{\sigma}_z|E_f^-\rangle)/(\rho_i+\rho_f)=2j/(\rho_i+\rho_f)$.
In the region ${\rm I\hspace{-.1em}I}$ ($x=0$), one particle arriving at $x=0$ from $x=\infty$ takes the velocity $T(E_f)c\langle\hat{\sigma}_z\rangle_{\bf w}^{\rm I\hspace{-.1em}I}$ in the average, of which process causes the steady fluxes in the regions I ($x<0$) and ${\rm I\hspace{-.1em}I\hspace{-.1em}I}$ ($x>0$).
Note that the reflection, which means holding the vacuum state, does not yield any net flux in the region ${\rm I\hspace{-.1em}I}$ ($x=0$).
Consequently, we expect the following equation,
\begin{eqnarray}
T(E_f)c\langle\hat{\sigma}_z\rangle_{\bf w}^{\rm I\hspace{-.1em}I} &=& \frac{2j}{\rho_i+\rho_f} \label{eq:trans_wv} \\
&=& 2c\left(\frac{1}{\langle E_i^+|\hat{\sigma}_z|E_i^+\rangle}+\frac{1}{\langle E_f^-|\hat{\sigma}_z|E_f^-\rangle} \right)^{-1}. \label{eq:trans_wv2}
\end{eqnarray}
In fact, equation (\ref{eq:trans_wv2}) gives us the very same $T(E_f)$ as equation (\ref{eq:trans}).

\begin{figure}
 \begin{center}
  \includegraphics[width=0.6\linewidth]{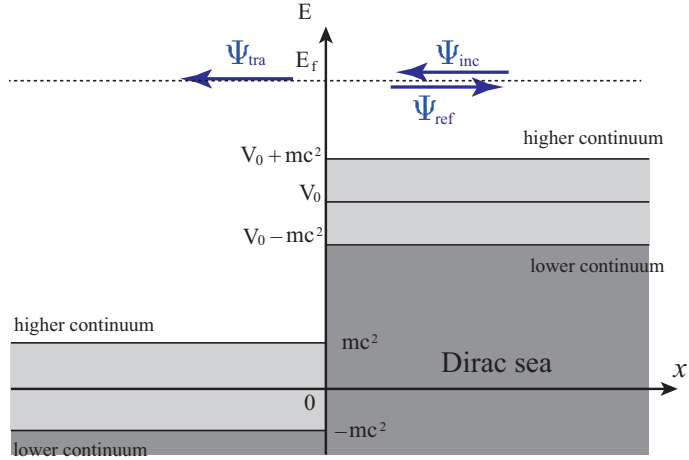}
  \caption{A particle with an energy being larger than $V_0+mc^2$ is incident from $x=\infty$.
The potential may satisfy $V_0<0$, although this figure shows the case of $V_0>0$.}
  \label{fig:tri_step}
 \end{center}
\end{figure}

Finally, we consider what value the weak value takes when a pair production can occur.
According to equation (\ref{eq:wv_vel}), the weak value is described as follows,
\begin{eqnarray}
\langle\hat{\sigma}_z\rangle_{\bf w}^{\rm I\hspace{-.1em}I}=\frac{E_i-E_f}{(p_i+p_f)c}=\frac{-V_0}{\sqrt{(E_f-V_0)^2-(mc^2)^2}+\sqrt{E_f^2-(mc^2)^2}}. \label{eq:str_wv}
\end{eqnarray}
We have assumed the conditions $E_f>mc^2$, and $V_0-mc^2>E_f$ for a pair production.
Under these conditions, the denominator of equation (\ref{eq:str_wv}) is positive and less than $\sqrt{(E_f-V_0)^2}+\sqrt{E_f^2}=V_0$.
Then, the weak value satisfies $\langle\hat{\sigma}_z\rangle_{\bf w}^{\rm I\hspace{-.1em}I}<-1$, which gives a strange weak value (superluminal velocity), because the average of $\hat{\sigma}_z$ satisfies $|\langle\hat{\sigma}_z\rangle|\le1$ conventionally.
On the other hand, as a trivial case, we can consider a transmission on an energy level from the higher continuum of $x>0$ to the higher continuum of $x<0$ as shown in figure \ref{fig:tri_step}.
In this case, the incident wave, which has the kinetic energy $E_i$ and the momentum $-p_i$, can transmit and result in the transmitted wave with $E_f$ and $-p_f$, where we have also used the definition of $p_i, p_f>0$ and $E_i=E_f-V_0$.
We need the conditions $E_f>mc^2$ and $V_0+mc^2<E_f$ in this case.
The above discussion for deriving the transmission probability (equations (\ref{eq:trans}) and (\ref{eq:trans_wv2})) and the weak value (equation (\ref{eq:wv_vel})) is available.
We can find the weak value as follows,
\begin{eqnarray}
\langle\hat{\sigma}_z\rangle_{\bf w}^{\rm I\hspace{-.1em}I}&=&\frac{E_i-E_f}{(-p_i+p_f)c}=\frac{-V_0}{-\sqrt{(E_f-V_0)^2-(mc^2)^2}+\sqrt{E_f^2-(mc^2)^2}} \nonumber \\
&=& \frac{-1}{2E_f-V_0}\left[\sqrt{(E_f-V_0)^2-(mc^2)^2}+\sqrt{E_f^2-(mc^2)^2}\right](<0) \nonumber \\
&>& \frac{-1}{2E_f-V_0}\left[\sqrt{(E_f-V_0)^2}+\sqrt{E_f^2}\right]=-1, \label{eq:tri_wv}
\end{eqnarray}
which satisfies $|\langle\hat{\sigma}_z\rangle_{\bf w}^{\rm I\hspace{-.1em}I}|\le1$.
Generally, we can find a strange weak value ($|\langle\hat{\sigma}_z\rangle_{\bf w}^{\rm I\hspace{-.1em}I}|>1$) when a transmission on an energy level is achieved from a higher (lower) continuum to a lower (higher) continuum, where the step potential is supercritical for such an energy level.
Meanwhile, a weak value is not strange ($|\langle\hat{\sigma}_z\rangle_{\bf w}^{\rm I\hspace{-.1em}I}|\le1$) for a transmission from a higher (lower) continuum to a higher (lower) continuum.

\section{Conclusion}
\label{sec:con}
We showed that a time evolution by $1+1$ dimensional space-time Dirac equation can be described with a weak value, when a system can be represented by a space part and a chirality separately.
As an application, we considered plane waves and spontaneous pair productions via a supercritical step potential.
With a weak value of a group velocity, we can explain the phenomenon reasonably.

A weak value often gives us a simple description for a physical phenomenon.
In relation to our work, we have noticed the paper which proposed an interpretation of a pair production as a result of weak measurement \cite{pair_pro}.
In \cite{pair_pro}, it was shown that the matrix elements can be understood as weak values on the quantitative analysis.
They also pointed out the significance of a weak value beyond its detectability.
In our case, due to the simple setup, we could find that a weak value can evidently be a value of a physical quantity:
We have shown an assured case in which a weak value is undoubtedly more than just a mathematical by-product.
Our case allows us to treat a quantum particle like a classical particle with a group velocity given by a weak value.
In addition, it has been revealed that the appearance of a strange weak value can be linked to a physical phenomenon, namely, a pair production.
We can find some recent works on superluminal velocity in the context of a weak value \cite{spl3, spl4}.
At a glance, one may doubt considering superlumial velocity, even if it satisfies causality.
Briefly speaking, such a strange weak value comes from quantum interference.
As we have mentioned, however, it should not be easily concluded that a weak value is just a by-product from calculation.
Although we have emphasized the significance of a weak value irrespective of weak measurement,
from the starting point of view, it is surely important that a weak value can be experimentally obtained by measurement.
Recently, weak measurement attracts attention as a tool for measurement of a quantum state without destroying it \cite{wf, wf2}.
Then, it may be an interesting question whether we should treat a quantum state as an substantial being: Can a wave function be beyond a calculation tool?

Our result supports the usefulness of a weak value as an actual value of a physical quantity, over and above just a result of weak measurement.
Although we have shown only one example so far, we hope our implication can be helpful to understand a weak value more profoundly and to give us another description for quantum mechanics.

\section*{Acknowledgements}
This work was supported by JSPS Research Fellowships for Young Scientists
and the Funding Program for World-Leading Innovative R \& D on Science and Technology (FIRST).

\end{document}